\documentclass[a4paper, twocolumn]{article}
\usepackage{graphicx}
\usepackage{amsmath}

\begin{document}

\title{Graph state basis for Pauli Channels }
\author{Xiao-yu Chen, Li-zhen Jiang \\
{\small {College of Information and Electronic Engineering, Zhejiang
Gongshang University, Hangzhou, 310018, China}}}
\date{}

\maketitle

\begin{abstract}
We introduce graph state basis diagonalization to calculate the coherent
information of a quantum code passing through a Pauli channel. The scheme is
5000 times faster than the best known one for some concatenated repetition
codes, providing us a practical constructive way of approaching the quantum
capacity of a Pauli channel. The calculation of the coherent information of
non-additive quantum code can also be greatly simplified in graph state
basis.

PACS number(s): 03.67.-a, 03.65.Ud,

Keyword(s): graph state basis; degenerate quantum code; quantum capacity.
\end{abstract}
\\

The quantum coding theorem for noisy channels \cite{Lloyd} \cite{Barnum1}
\cite{Barnum2} \cite{Devetak} states that the quantum capacity $Q(\mathcal{N}%
)$ of a channel $\mathcal{N}$ is given by regularized coherent information:
\begin{equation}
Q(\mathcal{N})=\lim_{n\rightarrow \infty }\frac 1n\max_{\rho _n}I(\rho _n,%
\mathcal{N}^{\otimes n}),
\end{equation}
the r.h.s term has long been known an upper bound for $Q(\mathcal{N})$,
which is the content of the converse coding theorem \cite{Barnum1} \cite
{Barnum2} . The direct coding theorem, stating that r.h.s tem is actually
attainable, has been strictly proven by Devetak \cite{Devetak}. The coherent
information $I(\rho ,\mathcal{N})$ of a state $\rho $ with respect to the
noise $\mathcal{N}$ is defined by
\begin{equation}
I(\rho ,\mathcal{N})=S(\mathcal{N(}\rho ))-S(\mathcal{I}_A\otimes \mathcal{N}%
(\Psi _{A\rho })),
\end{equation}
where $S(\varrho )=-tr\varrho \log _2\varrho $ is the von Neumann entropy, $%
\Psi _{A\rho }$ is a purification of $\rho $, and $\mathcal{I}_A$ is the
identity operation on the ancilla system $A$. The last term, $S(\mathcal{I}%
_A\otimes \mathcal{N}(\Psi _{A\rho }))$, is the entropy exchange $S_e(\rho ,%
\mathcal{N})$ of $\rho $ with respect to $\mathcal{N}$.

For a code state $\rho ^C$ of $n$ qubits, denote $\sigma ^C=\mathcal{N}%
^{\otimes n}\mathcal{(}\rho ^C),\sigma ^{AC}=\mathcal{I}_A\otimes \mathcal{N}%
^{\otimes n}(\left| \Psi ^{AC}\right\rangle \left\langle \Psi ^{AC}\right|
), $ where $\left| \Psi ^{AC}\right\rangle $ is the purification of $\rho ^C$%
, the coherent information of the state $\rho ^C$ with respect to the noise $%
\mathcal{N}^{\otimes n}$ per qubit will be $I^{C\mathcal{N}}=\frac
1n(S(\sigma ^C)-S(\sigma ^{AC})).$ Thus $I^{C\mathcal{N}}$ is the lower
bound of $Q(\mathcal{N})$ according to quantum noisy coding theorem. It is
known that the one-shot capacity $Q_1(\mathcal{N})=\max_\rho I(\rho ,%
\mathcal{N})$ is exactly the maximum rate achievable with a
non-degenerate code for Pauli channel $\mathcal{N}$. That
$Q(\mathcal{N})>Q_1(\mathcal{N})$ is then established by the
construction of a massively degenerate code, this was accomplished
in the work of \cite{Shor} \cite{DiVincenzo} for depolarizing
channel and \cite{Smith} \cite{Fern} for some Pauli channels. It
is not known which quantum code achieves the quantum capacity for
a channel that is neither degradable nor anti-degradable. Pauli
channel with proper channel parameters is an example of such
channels. So we need to check all possible codings to seek the
maximal coherent information, this is an awful work in the
viewpoint of just working out the quantum capacity. However, the
history of classical communication tells us that coding is the
really important thing even when the capacity is known. The aim of
this paper is to provide a scheme to work out the coherent
information for a quantum code with respect to a Pauli channel.

\textit{Graph state basis}.--- A graph $G=(V;\Gamma )$ is composed of a set $%
V$ of $n$ vertices and a set of edges specified by the adjacency matrix $%
\Gamma $, which is an $n\times n$ symmetric matrix with vanishing diagonal
entries and $\Gamma _{ab}$ $=1$ if vertices $a,b$ are connected and $\Gamma
_{ab}$ $=0$ otherwise. The neighborhood of a vertex $a$ is denoted by $N_a$ $%
=\{v\in V\left| \Gamma _{av}=1\right. \}$, i.e, the set of all the vertices
that are connected to $a$. Graph states \cite{Hein1} \cite{Sch} are useful
multipartite entangled states that are essential resources for the one-way
computing \cite{Raus} and can be experimentally demonstrated \cite{Walther}.
To associate the graph state to the underlying graph, we assign each vertex
with a qubit, each edge represents the interaction between the corresponding
two qubits. More physically, the interaction may be Ising interaction of
spin qubits. Let us denote the Pauli matrices at the qubit $a$ by $%
X_a,Y_a,Z_a$ and identity by $I_a$. The graph state related to graph $G$ is
defined as
\begin{equation}
\left| G\right\rangle =\prod_{\Gamma _{ab}=1}U_{ab}\left| +\right\rangle
^V=\frac 1{\sqrt{2^n}}\sum_{\mathbf{\mu }=\mathbf{0}}^{\mathbf{1}%
}(-1)^{\frac 12\mathbf{\mu }\Gamma \mathbf{\mu }^T}\left| \mathbf{\mu }%
\right\rangle  \label{wave0}
\end{equation}
where $\left| \mathbf{\mu }\right\rangle $ is the joint eigenstate of Pauli
operators $Z_a$ ($a\in V$) with eigenvalues $(-1)^{\mu _a}$, $\left|
+\right\rangle ^V$ is the joint +1 eigenstate of Pauli operators $X_a$ ( $%
a\in V$) , and $U_{ab}$ ($U_{ab}=diag\{1,1,1,-1\}$ in the $Z$ basis) is the
controlled phase gate between qubits $a$ and $b$. Graph state can also be
viewed as the result of successively performing 2-qubit Control-Z operations
$U_{ab}$ to the initially unconnected $n$ qubit state $\left| +\right\rangle
_x^V$. It can be shown that graph state is the joint $+1$ eigenstate of the $%
n$ vertices stabilizers
\begin{equation}
K_a=X_a\prod_{b\in N_a}Z_b:=X_aZ_{N_a},\text{ }a\in V.
\end{equation}
Meanwhile, the graph state basis are $\left| G_{k_1,k_2,\cdots
k_n}\right\rangle $ $=\prod_{a\in V}Z_a^{k_a}\left| G\right\rangle ,$ with $%
k_a=0,1.$ Thus
\begin{equation}
K_a\left| G_{k_1,k_2,\cdots k_n}\right\rangle =(-1)^{k_a}\left|
G_{k_1,k_2,\cdots k_n}\right\rangle .
\end{equation}

\textit{Output state in graph state basis.--- }Consider the input code state
$\rho ^C$which is diagonal in graph state basis, that is,
\begin{equation}
\rho ^C=\sum_{k=0}^{2^n-1}\pi _k\left| G_k\right\rangle \left\langle
G_k\right| ,
\end{equation}
with $\pi _k$ the probability of state the graph state $\left|
G_k\right\rangle $ , $0\leq \pi _k\leq 1$ and $\sum_k\pi _k=1,$ we have
denoted $\left| G_{k_1,k_2,\cdots k_n}\right\rangle =$ $\left|
G_k\right\rangle $ with the conventions of $k=\sum_ik_i2^{n-i}$ and
\begin{equation}
k\oplus j=\sum_i(k_i\oplus j_i)2^{n-i}.  \label{wave1}
\end{equation}
The purification state can be $\left| \Psi ^{AC}\right\rangle =\sum_k\sqrt{%
\pi _k}\left| G_k\right\rangle _A\left| G_k\right\rangle .$ In Krauss
representation, Pauli channel map $\mathcal{N}$ acting on qubit state $\rho $
can be written as $\mathcal{N}\left( \rho \right) =f\rho +p_xX\rho
X+p_yY\rho Y+p_zZ\rho Z,$ where $p_{x(y,z)}\in [0,1]$ are the probabilities,
$f=1-p_x-p_y-p_z$ $\in [0,1]$ is the fidelity of the channel, and $X,Y,Z$
are the Pauli operators. For depolarizing channel, $p_x=p_y=p_z=p,$ $f=1-3p.$
For $n$ use of channels with $n$ qubits input state $\rho _n$, we have the
output state $\mathcal{N}^{\otimes n}\left( \rho _n\right) =\sum_a\eta
_aE_a\rho _nE_a^{\dagger },$ with $\eta _a=f^{n-i-j-l}p_x^ip_y^jp_z^l$ for $%
E_a=X^iY^jZ^l.$ Then the joint output state of $\rho ^C$ and the ancilla is $%
\sigma ^{AC}=\sum_{ij}\sqrt{\pi _i\pi _j}(\left| G_i\right\rangle
_A\left\langle G_j\right| _A)$ $\otimes \mathcal{N}^{\otimes n}(\left|
G_i\right\rangle \left\langle G_j\right| )$ $=\sum_{ij}\sqrt{\pi _i\pi _j}%
\sum_a\eta _a(\left| G_i\right\rangle _A\left\langle G_j\right| _A)$ $%
\otimes (E_a\left| G_i\right\rangle \left\langle G_j\right| E_a^{\dagger }).$
In graph state basis, we have $\sigma _{sm,tl}^{AC}=\left\langle G_s\right|
_A\left\langle G_m\right| \sigma ^{AC}\left| G_t\right\rangle _A\left|
G_l\right\rangle $ $=\sum_{ij}\sqrt{\pi _i\pi _j}\sum_a\eta _a\left\langle
G_m\right| E_a\left| G_i\right\rangle \left\langle G_j\right| E_a^{\dagger
}\left| G_l\right\rangle \delta _{si}\delta _{tj}.$ Thus the matrix elements
of the joint output state in graph state basis are
\begin{eqnarray}
\sigma _{im,jl}^{AC} &=&\sqrt{\pi _i\pi _j}\sum_a\eta _a\left\langle
G_m\right| E_a\left| G_i\right\rangle \left\langle G_j\right| E_a^{\dagger
}\left| G_l\right\rangle  \nonumber \\
&=&\sqrt{\pi _i\pi _j}\sum_a\eta _a\left\langle G\right|
Z^{(m)}E_aZ^{(i)}\left| G\right\rangle  \nonumber \\
&&\times \left\langle G\right| Z^{(j)}E_a^{\dagger }Z^{(l)}\left|
G\right\rangle ,
\end{eqnarray}
where we have denoted $Z^{(k)}=\prod_{c\in V}Z_c^{k_c}.$ According
to the orthogonality of graph state basis, $\left\langle G\right|
Z^{(m)}E_aZ^{(i)}\left| G\right\rangle =0$ except
$Z^{(m)}E_aZ^{(i)}=K_s$ up to a factor of $\pm 1,\pm i$ (the
factor will be omit hereafter because it has no effect in the
elements of $\sigma ^{AC}$) $,$ for some $K_s\in K$ (the group
with generators of all $K_c,c\in V,$ an Abelian group, the
vertices stabilizer group)$,.$ Thus we have $E_a=Z^{(m)}K_sZ^{(i)},$ so $%
Z^{(j)}E_a^{\dagger }Z^{(l)}=Z^{(j)}Z^{(i)}K_sZ^{(m)}Z^{(l)}.$ It can be
written as $Z^{(i\oplus j)}K_sZ^{(m\oplus l)}$ by our convention (\ref{wave1}%
). Since $Z^{(i\oplus j)}$ either commutates or anti-commutates with $K_s,$
we have $Z^{(j)}E_a^{\dagger }Z^{(l)}=\pm K_sZ^{(i\oplus j\oplus m\oplus
l)}. $ For a non-zero $\left\langle G\right| Z^{(j)}E_a^{\dagger
}Z^{(l)}\left| G\right\rangle ,$ we should get
\begin{equation}
i\oplus j\oplus m\oplus l=0.
\end{equation}
Let $k=i\oplus m,$ then $l=i\oplus j\oplus m=j\oplus k,$and $m=i\oplus k$ so
the possible non-zero elements are in the form of
\begin{equation}
\sigma _{i(i\oplus k),j(j\oplus k)}^{AC}=\sqrt{\pi _i\pi _j}\sum_{a\left|
E_a\in Z^{(k)}K\right. }(-1)^{P_a}\eta _a,  \label{wave2}
\end{equation}
where $P_a=0$ with $E_a=Z^{(i+k)}K_sZ^{(i)}$ such that $Z^{(i\oplus j)}$
communicates with $K_s,$and $P_a=1$ with $E_a=Z^{(i+k)}K_sZ^{(i)}$such that $%
Z^{(i\oplus j)}$ anti-communicates with $K_s.$ Notice that the
joint output state can be block diagonalized according to $k$. So
that what left is to diagonaling each block with give $k$. The
routine of calculating the non-zero elements is like this: (1) to
list all the elements of $K,$ (2) to determine $P$ according to
the commutator of $Z^{(i\oplus j)}$and $K,$(3) to multiply
$Z^{(k)}$ to obtain the coset $E=Z^{(k)}K$ of the Pauli group with
respect to its subgroup $K$ and to determine $\eta $,

Notice that for $i^{\prime }\oplus j^{\prime }=i\oplus j,$ we have
$\sigma _{i(i\oplus k),j(j\oplus k)}^{AC}/\sqrt{\pi _i\pi
_j}=\sigma _{i^{\prime }(i^{\prime }\oplus k),j^{\prime
}(j^{\prime }\oplus k)}^{AC}/\sqrt{\pi _i\pi _j}.$ This property
is very useful in further diagonalizing the submatrix for
stabilizer code of equal probability.

\textit{Stabilizer code.--- }In a graphical quantum error-correction code,
each codeword can be written as $\left| G_k\right\rangle =Z^{(k)}\left|
G\right\rangle =\prod_{c\in V}Z_c^{k_c}\left| G\right\rangle $. To encode is
to properly choose some of the $\left| G_k\right\rangle $ in order to form
the code. A code is thus completely characterized by the set of $k$ for a
underneath given graph. For stabilizer code encoding $q$ qubits into $n$
qubits, all the $2^q$ chosen $Z^{(k)}$ forms a.group, each $Z^{(k)}$ is self
inverse. Without loss of generality, we use the binary vector $(\overline{k_1%
},\overline{k_2},\ldots ,\overline{k_q})$ to characterize the
codeword of the stabilizer code. Then quantum stabilizer encoding
is an encoding of
classical binary serial $(\overline{k_1},\overline{k_2},\ldots ,\overline{k_q%
})$ into binary serial $(k_1,k_2,\ldots ,k_n).$ Denote $\overline{k}%
=\sum_{l=1}^q\overline{k_l}2^{l-1},$ we have $\overline{i^{\prime }}\oplus
\overline{j^{\prime }}=\overline{i}\oplus \overline{j}$ if $i^{\prime
}\oplus j^{\prime }=i\oplus j,$ when $i,j,i^{\prime },j^{\prime }$
correspond to codewords. A matrix $M$ with $M_{\overline{i}\overline{j}}=M_{%
\overline{i}\oplus \overline{l},\overline{j}\oplus \overline{l}},$ can be
diagonalized with Hadamard matrix $H_q$. Its eigenvalues will be $\sqrt{2^q}%
(H_qM)_{\overline{i},\overline{0}}.$ Each block of $\sigma ^{AC}$
can be written in the form of $M\ $for a code with a priori equal
probability $\pi _{\overline{i}}=2^{-q}.$ Hence, the eigenvalues
of each bloc $k$ of $\sigma ^{AC}$ will be
\begin{equation}
\lambda _{\overline{i}}=\frac 1{\sqrt{2^q}}\left( H_qM\right) _{\overline{i},%
\overline{0}}.
\end{equation}
The channel output state is simply $\sigma ^C=Tr_A\sigma ^{AC}$, the matrix
element of $\sigma ^C$ in graph state basis is
\begin{equation}
\sigma _{km}^C=\sum_i\sigma _{i(i\oplus k),i(i\oplus k)}^{AC}\delta _{km}
\label{wave3}
\end{equation}
In graph state basis, the output state of a Pauli channel with diagonal
input is still diagonal, as mentioned in \cite{Acin}. Thus far, we have
obtained all the eigenvalues for calculating the coherent information of the
input of a priori uniform distributed stabilizer code with respect to Pauli
channel.

\textit{Concatenated repetition codes in depolarizing channel.---}%
Depolarizing channel is a special case of Pauli channel, we will
mainly deal with the depolarizing channel, the results can easily
be extended to generic Pauli channel.\textit{\ }One way to show
$Q(\mathcal{N})>Q_1(\mathcal{N})$
is to find $Q(\mathcal{N})>0$ for very noisy channel $\mathcal{N}$ where $%
Q_1(\mathcal{N})=0.$ Some codes that were shown to allow correction in the
range of $Q_1(\mathcal{N})=0$ consist of an $n_1$ qubit bit flip code
concatenated with an $n_2$ qubit phase flip code \cite{Smith} \cite
{DiVincenzo} \cite{Fern}. These have been called ''$n_1$ in $n_2$''codes
\cite{Fern}, since each of the $n_2$ blocks of the phase flip code consists
of an $n_1$ qubit bit flip code. One of the examples is the famous Shor $%
[[9,1,3]]$ code which is the ''3 in 3'' code. The codewords of ''$n_1$ in $%
n_2$''code can be obtained similarly as in Shor $[[9,1,3]]$ code, they are
\begin{eqnarray}
\left| \overline{0}\right\rangle &=&\frac 1{\sqrt{2^{n_1n_2}}}(\left|
0^{\otimes n_1}\right\rangle +\left| 1^{\otimes n_1}\right\rangle )^{\otimes
n_2}, \\
\left| \overline{1}\right\rangle &=&\frac 1{\sqrt{2^{n_1n_2}}}(\left|
0^{\otimes n_1}\right\rangle -\left| 1^{\otimes n_1}\right\rangle )^{\otimes
n_2},
\end{eqnarray}
where $\frac 1{\sqrt{2^{n_1}}}(\left| 0^{\otimes n_1}\right\rangle
\pm \left| 1^{\otimes n_1}\right\rangle )$ are the $n_1-$partite
GHZ states. The ''$n_1$ in $n_2$''codeword is the repetition of
GHZ state. Comparing with the definition (\ref{wave0}) of graph
state, we find that the underneath graph for the code can be the
''forest'' graph. The graph contains $n_2$ independent and
identical subgraphs, each subgraph of $n_1$ vertex has a tree
structure with the root vertex connecting with all the other
vertices and no other links exist. The vertices will be numbered
in the following order: the $j_1-th$ leaf of the $j_2-th$ tree is
numbered as the $(j_2(n_1-1)+j_1)-th$ vertex, the root of $j_2-th$
tree is numbered as the $j_2n_1-th$ vertex. We have $\left|
\overline{0}\right\rangle =\left| G\right\rangle $ and $\left|
\overline{1}\right\rangle =Z_{n_1}Z_{2n_1}\cdots Z_{n_2n_1}\left|
G\right\rangle $ , where $\left| G\right\rangle $ is the graph
state of the ''forest'' graph. The input state to the channel is
$\rho ^C=\frac 12(\left| \overline{0}\right\rangle \left\langle
\overline{0}\right| +\left| \overline{1}\right\rangle \left\langle
\overline{1}\right| ),$ where the equal a priori probabilities are
assumed.

To simplify the analysis, we consider the case of $n_2=1,$ the tree graph
first. Now $\left| \overline{0}\right\rangle =\left| G\right\rangle $ and $%
\left| \overline{1}\right\rangle =Z_{n_1}\left| G\right\rangle .$ The
vertices stabilizer group $K$ can be divided into its subgroup $K^e$ with
generators $K_1,\ldots ,K_{n_1-1}$ and the coset $K^o=K_{n_1}K^e,$ so that
all the elements of $K^e$ commutate with $Z_{n_1}$ and all the elements of $%
K^o$ anti-commutate with $Z_{n_1}.$ The index $i$ and $j$ can be
$(0,\cdots
,0,0)$ or $(0,\cdots ,0,1).$ Correspondingly, $\overline{i}$ and $\overline{j%
}$ can be $\overline{0}$ or $\overline{1}$. In the basis of $\left|
\overline{0}\right\rangle $ and $\left| \overline{1}\right\rangle ,$each
block of the joint output state $\sigma ^{AC}$ is a $2\times 2$ matrix in
the form of
\begin{equation}
\frac 12\left[
\begin{array}{ll}
\eta _e+\eta _o & \eta _e-\eta _o \\
\eta _e-\eta _o & \eta _e+\eta _o
\end{array}
\right] ,  \label{wave4}
\end{equation}
where $\eta _e=\sum_a\eta _a$ with the condition of $E_a\in Z^{(k)}K^e;$ $%
\eta _o=\sum_a\eta _a$ with the condition of $E_a\in Z^{(k)}K^o\ $for the $%
k-th$ block. For each block, $\eta _e$ and $\eta _o$ are the eigenvalues of $%
\sigma ^{AC}$.

For the $k=0$ block of the joint output density matrix in Eq. (\ref{wave2}),
we have $E=K.$ Now $K^e$ is generated by $K_l=X_lZ_{n_1}$ $(l=1,\cdots
,n_1-1),$ the group elements of $K^e$ can be classified as two classes: the
even class and the odd class. The even class have even generators, the group
elements will be $I,X_1X_2,\cdots X_{n_1-2}X_{n_1-1},\cdots ,$ the
contribution to $\eta _e$ is $f^{n_1}+\binom{n_1-1}2f^{n_1-2}p^2+$ $\binom{%
n_1-1}4f^{n_1-4}p^4+\cdots =\frac 12[(f+p)^{n_1-1}+(f-p)^{n_1-1}]f.$ The odd
class have odd generators, the group elements will be $%
X_1Z_{n_1},X_2Z_{n_1},\cdots ,X_{n_1-1}Z_{n_1},X_1X_2X_3Z_{n_1}\cdots ,$ the
contribution to $\eta _e$ is $\binom{n_1-1}1f^{n_1-2}p^2$ $+\binom{n_1-1}%
3f^{n_1-4}p^4$ $+$ $\cdots $ $=\frac 12[(f+p)^{n_1-1}-(f-p)^{n_1-1}]p.$
Denote $x=p/f=p/(1-3p),$we have $\eta _e(k=0)$ $=\frac{f^{n_1}}%
2[(1+x)^{n_1}+(1-x)^{n_1}].$ While for $\eta _o$, since $%
K_{n_1}=X_{n_1}Z_1Z_2\cdots Z_{n_1-1},$the number of the Pauli error
operators in each element of $K^o=K_{n_1}K^e$ is $n_1,$ so $\eta _o(k=0)$ $%
=2^{n_1-1}p^{n_1}=f^{n_1}2^{n_1-1}x^{n_1}.$

The next block we should consider is with the coset $E=Z_1K,$with $%
E^e=Z_1K^e $ and $E^o=Z_1K^o.$ The group $K^e$ now can be classified as two
class: the class without $K_1$ and the class with $K_1.$ The class without $%
K_1$ contributes to $\eta _e$ with $\frac{f^{n_1}}%
2[(1+x)^{n_1-1}+(1-x)^{n_1-1}]x, $the last factor $x$ comes from $Z_1$ in $%
E^e=Z_1K^e.$ The class with $K_1$ contributes to $\eta _e$ with $\frac{%
f^{n_1}}2[(1+x)^{n_1-1}-(1-x)^{n_1-1}]x, $ where the set $%
Z_1K_1\{I,X_2X_3,\cdots \}$ contributes $\frac{f^{n_1}}%
2[(1+x)^{n_1-2}+(1-x)^{n_1-2}]x^2$ and the set $Z_1K_1%
\{X_2Z_{n_1},X_3Z_{n_1},\cdots \}$ contributes $\frac{f^{n_1}}%
2[(1+x)^{n_1-2}-(1-x)^{n_1-2}]x.$ Thus we have $\eta _e(k=2^{n_1-1})=$ $%
f^{n_1}(1+x)^{n_1-1}x.$ While for $\eta _o,$ we have $%
E^o=Z_1K^o=Z_1K_{n_1}K^e.$ The class without $K_1$ contributes to $\eta _o$
with $f^{n_1}2^{n_1-2}x^{n_1-1},$the class with $K_1$ contributes to $\eta
_o $ with $f^{n_1}2^{n_1-2}x^{n_1},$ hence $\eta _o(k=2^{n_1-1})$ $%
=f^{n_1}2^{n_1-2}(1+x)x^{n_1-1}.$ For all the cosets $E=Z_lK$ $(l=2,\ldots
,n_1-1),$ we obtain the same results of $\eta _e$ and $\eta _o$ as in the
case of the coset $Z_1K.$ A further research shows that for coset $%
Z_{l_1}Z_{l_2}\cdots Z_{l_m}K$ with $1\leq l_i\leq n_1-1,$ we can get $\eta
_e=f^{n_1}2^{m-1}(1+x)^{n_1-m}x^m,\eta
_o=f^{n_1}2^{n_1-m-1}(1+x)^mx^{n_1-m}. $

For the coset $E=Z_{n_1}K,$ we can get $\eta _e(k=1)$ $=\frac{f^{n_1}}%
2[(1+x)^{n_1}-(1-x)^{n_1}],\eta _o(k=1)$
$=f^{n_1}2^{n_1-1}x^{n_1}.$ Similarly, for coset
$Z_{l_1}Z_{l_2}\cdots Z_{l_m}Z_{n_1}K$ with $1\leq l_i\leq n_1-1,$
we have $\eta _e(k=1)=t_{m+1},\eta _o(k=1)=t_{n_1-m+1}.$ The
eigenvalues of the joint output state $\sigma ^{AC}$ of tree graph
case are
\begin{eqnarray}
t_0 &=&\frac{f^{n_1}}2[(1+x)^{n_1}+(1-x)^{n_1}], \\
t_1 &=&\frac{f^{n_1}}2[(1+x)^{n_1}-(1-x)^{n_1}], \\
t_{m+1} &=&f^{n_1}2^{m-1}(1+x)^{n_1-m}x^m,\text{ }1\leq m\leq n_1
\end{eqnarray}
The eigenvalues of $\sigma ^{AC}$ whose value is $t_{m+1}(1\leq m\leq n_1-1)$
has the degeneracy $2\binom{n_1-1}m=2C_{n_1-1}^m.$ From Eq. (\ref{wave3}),
the eigenvalues of the output state $\sigma ^C$ are $\frac 12(\eta
_e(k)+\eta _o(k))$ $+\frac 12(\eta _e(k+1)+\eta _o(k+1)),$ which are $\frac
12(t_0+t_1+2t_{n_1+1})\ $and $\frac 12(t_{m+1}+t_{n_1-m+1})$ for $(1\leq
m\leq n_1-1).$

Denote the ''forest'' vertices stabilizer group as $\mathcal{K},$with its
subgroup $\mathcal{K}_{j_2}$ for $j_2-th$ tree$.$ Let's split the ''forest''
vertices stabilizer group $\mathcal{K}$ into two parts according to the
commutators of the elements and $Z_{n_1}Z_{2n_1}\cdots Z_{n_2n_1}.$ The
commutator of the element of $\mathcal{K}$ and $Z_{n_1}Z_{2n_1}\cdots
Z_{n_2n_1}$ for ''forest'' graph can be reduced to the product the
commutators of the corresponding piece of $\mathcal{K}_{j_2}$of and $%
Z_{j_2n_1}$. When the element of $\mathcal{K}$ commutates with $%
Z_{n_1}Z_{2n_1}\cdots Z_{n_2n_1}$ , the number of the trees with
anti-commutator of its corresponding section of the element of $\mathcal{K}%
_{j_2}$ and $Z_{j_2n_1}$ should be even. When the element of $\mathcal{K}$
anti-commutates with $Z_{n_1}Z_{2n_1}\cdots Z_{n_2n_1}$, the number of that
should be odd. For a given coset $\mathcal{E}=\mathcal{Z}^{(k)}\mathcal{K}$
of the ''forest''$,$ the ''forest'' coset head $\mathcal{Z}^{(k)}$ is
composed of $n_2$ sections, each section is the coset head of the tree graph
case. Since the trees are identical, the types of the sections can be
denoted with $I,Z_{n_1},Z_1,$ $Z_1Z_{n_1},Z_1Z_2,$ $Z_1Z_2Z_{n_1},$ $\cdots
,Z_1Z_2\cdots Z_{n_1-1},$ $Z_1Z_2\cdots Z_{n_1-1}Z_{n_1}.$ Here type $%
Z_1Z_2\cdots Z_m$ $(1\leq m\leq n_1-1)$ represents all the cosets whose $Z$
operator numbers are $m.$ While the eigenvalues of the cosets $Z_1Z_2\cdots
Z_m$ and $Z_1Z_2\cdots Z_mZ_{n_1}$ are equal, we can further simplify the
types of coset head sections as $I,Z_{n_1},Z_1,$ $Z_1Z_2,$ $\cdots
,Z_1Z_2\cdots Z_{n_1-1},$ with degeneracies $1,1,$ $2C_{n_1-1}^1,$ $%
2C_{n_1-1}^2\cdots $ $2C_{n_1-1}^{n_1-1}$, respectively.

For a given coset head $\mathcal{Z}$, suppose the number of trees with coset
head $I$ is $l_0,$the number of trees with coset head type $Z_{n_1}$ is $%
l_1, $the number of trees with coset head type $Z_1Z_2\cdots Z_m$ is $%
l_{m+1}.$ The total number of the trees is $n_2=\sum_{m=-1}^{n_1}l_{m+1}.$
For a particular element of $\mathcal{K},$ consider the type $Z_1Z_2\cdots
Z_m$ trees, suppose there be $s_{m+1}$ of the trees with their corresponding
piece of $\mathcal{K}_{j_2}$ anti-commutating with their own $Z_{j_2n_1}$,
this element of $\mathcal{K}$ should contribute to the eigenvalues of the
joint output state of the concatenated code with
\begin{equation}
t_0^{l_0-s_0}t_1^{l_1-s_1}t_{n_1+1}^{s_0+s_1}%
\prod_{m=1}^{n_1-1}t_{m+1}^{l_{m+1}-s_{m+1}}t_{n_1-m+1}^{s_{m+1}}
\end{equation}
Summing upon all elements of $\mathcal{K},$ we arrive at
\begin{eqnarray}
\eta ^e+\eta ^o &=&(t_0+t_{n_1+1})^{l_0}(t_1+t_{n_1+1})^{l_1}  \nonumber \\
&&\times \prod_{m=1}^{n_1-1}(t_{m+1}+t_{n_1-m+1})^{l_{m+1}},  \label{wave5}
\\
\eta ^e-\eta ^o &=&(t_0-t_{n_1+1})^{l_0}(t_1-t_{n_1+1})^{l_1}  \nonumber \\
&&\times \prod_{m=1}^{n_1-1}(t_{m+1}-t_{n_1-m+1})^{l_{m+1}},  \label{wave6}
\end{eqnarray}
Denote $a_{0\pm }=t_0\pm t_{n_1+1},a_{1\pm }=t_1\pm t_{n_1+1},a_{(m+1)\pm
}=t_{m+1}\pm t_{n_1-m+1}$ $(2\leq m\leq n_1-1),$ notice that $a_{(m+1)\pm
}=\pm a_{(n_1-m+1)\pm }$ for $2\leq m\leq n_1-1,$ we may only use $a_{m+1}$
with $1\leq m\leq \left\lceil \frac{n_1-1}2\right\rceil $ to specify the
eigenvalues besides $a_{0\pm },a_{1\pm }$. Consider the factors $%
(t_{m+1}+t_{n_1-m+1})^{l_{m+1}}$ and $(t_{n_1-m+1}+t_{m+1})^{l_{n_1-m+1}}$
in Eq. (\ref{wave5}), their degeneracies are $(2C_{n_1-1}^m)^{l_{m+1}}$ and $%
(2C_{n_1-1}^{n_{1-}m})^{l_{n_1-m+1}}$, respectively. For $1\leq m\leq
\left\lfloor \frac{n_1-1}2\right\rfloor $ , let $\mu
_{m+1}=l_{m+1}+l_{n_1-m+1},$ the degeneracy of $a_{(m+1)+}$ should be $%
\sum_{l_{m+1}=0}^{\mu _{m+1}}C_{\mu
_{m+1}}^{l_{m+1}}(2C_{n_1-1}^m)^{l_{m+1}}(2C_{n_1-1}^{n_{1-}m})^{\mu
_{m+1}-l_{m+1}}=(2C_{n_1-1}^m+2C_{n_1-1}^{n_{1-}m})^{\mu
_{m+1}}=(2C_{n_1}^m)^{\mu _{m+1}}\ .$ For even $n_1$ and $m=\frac{n_1}2,$let
$\mu _{m+1}=l_{m+1}$ the degeneracy of $a_{(m+1)+}$ $=2t_{m+1}$will be $%
(2C_{n_1-1}^m)^{\mu _{m+1}}=(C_{n_1}^m)^{\mu _{m+1}}.$ Thus the number of
coset that gives the same $\eta ^e+\eta ^o$ is
\begin{equation}
d\left( \mu \right) =2^hn_2!\prod_{m=1}^{\left\lceil (n_1-1)/2\right\rceil
}\frac 1{\mu _{m+1}!}(C_{n_1}^m)^{\mu _{m+1}},  \label{wave7}
\end{equation}
where vector $\mu =(\mu _0,\mu _1,\cdots ,\mu _{\left\lceil
(n_1-2)/2\right\rceil }),$ $h=\sum_{m=1}^{\left\lfloor
(n_1-1)/2\right\rfloor }\mu _{m+1},$ with $\sum_{m=-1}^{\left\lceil
(n_1-1)/2\right\rceil }\mu _{m+1}=n_2$ and $\mu _{0,1}=l_{0,1}.$

Let $n_3=\left\lceil (n_1-2)/2\right\rceil +1,$ the eigenvalues of the joint
output state $\sigma ^{AC}$ of ''$n_1$ in $n_2$''code can be written as
\begin{equation}
\eta _{\pm }\left( \mu \right) =\frac 12(\prod_{m=0}^{n_3}a_{m+}^{\mu _m}\pm
\prod_{m=0}^{n_3}a_{m-}^{\mu _m}),  \label{wave8}
\end{equation}
with degeneracy $d\left( \mu \right) $. According to Eq.(\ref{wave3}), the
eigenvalues of the output state $\sigma ^C$ are
\begin{equation}
\eta ^{\prime }\left( \mu \right) =\frac 12(a_{0+}^{\mu _0}a_{1+}^{\mu
_1}+a_{1+}^{\mu _0}a_{0+}^{\mu _1})\prod_{m=2}^{n_3}a_{m+}^{\mu _m},
\label{wave9}
\end{equation}
also with degeneracy $d\left( \mu \right) .$ The coherent information of ''7 in $%
n_2"$ code per channel use should be
\begin{eqnarray}
I^{C\mathcal{N}} &=&\frac 1{n_1n_2}\sum_\mu d\left( \mu \right) [-\eta
^{\prime }\left( \mu \right) \log _2\eta ^{\prime }\left( \mu \right)
\nonumber \\
&&+\eta _{+}\left( \mu \right) \log _2\eta _{+}\left( \mu \right) +\eta
_{-}\left( \mu \right) \log _2\eta _{-}\left( \mu \right) .  \label{wave10}
\end{eqnarray}

\textit{Examples.--- }Consider ''7 in $n_2"$ code, we have
$d\left( \mu \right) =\frac{n_2!}{\mu _2!\mu _3!\mu _4!}14^{\mu
_2}42^{\mu _3}70^{\mu _4}, $ $a_{0\pm }=\frac
12[(1-2p)^7+(1-4p)^7\pm (2p)^7],$ $a_{1\pm }=\frac
12[(1-2p)^7-(1-4p)^7\pm (2p)^7],$ $a_{2\pm }=\frac
12(1-2p)^2(1-4p)^2[(1-2p)^5\pm (1-4p)^5],$ $a_{3\pm }=\frac
12(1-2p)^4(1-4p)^4[(1-2p)^3\pm (1-4p)^3],$ $a_{4\pm }=\frac
12(1-2p)^6(1-4p)^6[(1-2p)\pm (1-4p)].$ The coherent information
can be quickly calculated. The typical time to obtain the optimal
$n_2=133$ is an hour (2004 CPU), while the former best result
needs a week (2008 CPU)\cite {Fern}. Further quick calculation is
also possible by expanding the logarithmic in Eq.(\ref{wave10}),
the time required for obtaining the optimal $n_2$ for ''7 in
$n_2"$ code is 2 minutes. For generic ''$n_1$ in $n_2"$ code we
have
\[
I^{C\mathcal{N}}=\frac 1{n_1n_2}(v_0-v_1+v_2-1),
\]
with
\begin{eqnarray*}
v_0 &=&n_2(a_{0+}\log _2a_{0+}+a_{1+}\log _2a_{1+}), \\
v_1 &=&\sum_{\mu _0,\mu _1}\frac{n_2!(1-a_{0+}-a_{1+})^{n_2-\mu _0-\mu _1}}{%
\mu _0!\mu _1!(n_2-\mu _0-\mu _1)}c(\mu _0,\mu _1), \\
v_2 &=&\frac 1{\ln 2}\sum_{l=0}^\infty \frac 1{2l(2l-1)}\left[
\sum_{j=0}^{\left\lceil (n_2-1)/2\right\rceil
}w_ja_{j+}(b_{j+}/a_{j+})^{2l}\right] ^{n_2},
\end{eqnarray*}
where $c(\mu _0,\mu _1)=\frac 12(a_{0+}^{\mu _0}a_{1+}^{\mu _1}+a_{1+}^{\mu
_0}a_{0+}^{\mu _1})\log _2[\frac 12(a_{0+}^{\mu _0}a_{1+}^{\mu
_1}+a_{1+}^{\mu _0}a_{0+}^{\mu _1})],$ $w_0=1,$ $w_1=1,$ $%
w_j=2C_{n_1}^j(2\leq j\leq \left\lfloor (n_2-1)/2\right\rfloor ,$ $%
w_{n_1/2}=C_{n_1}^{n_1/2}$ (for even $n_1$). New results of optimal $n_2$
are list in Table 1,
\begin{table}[tbp]
{\bfseries Table 1}\\[1ex]
\begin{tabular}{p{15ex}p{15ex}p{15ex}}
\hline
$n_1$ & $n_2$ & $3p_{\max }$ \\ \hline
9 & 334 & 0.19080163947 \\
10 & 113 & 0.19005842449 \\
11 & 828 & 0.19047624366 \\
12 & 309 & 0.18981106971 \\
14 & 812 & 0.18953710664 \\ \hline
\end{tabular}
\end{table}
with $p_{\max }$ being the critical value of channel noise such that $I^{C%
\mathcal{N}}(p_{\max })=0,$and for $p<p_{\max }$ we have positive coherent
information. The ''$13$ in $n_2"$ code has an optimal $n_2>1020.$ Too large $%
n_2$ makes the storage of $C_{n_2}^{\left\lfloor
n_2/2\right\rfloor }$ overflow.

The density matrix of the joint output can also be block
diagonalizable for a non-additive code input which is diagonal in
graph state basis. For example, the eigenvalues of the output and
joint output of $((5,6,2))$ code with respect to depolarizing
channel can be obtained analytically when the input is diagonal
and with equal probability in graph state basis. The density
matrix of the joint output can be block diagonalized as $32$
blocks, each block is a $6\times 6$ matrix and can be diagonalized
eventually.

In summary, we have block diagonalized the output density matrix
of a code and the joint output density of the code and the ancilla
system with respect to Pauli channel when the input quantum code
is diagonal in graph state basis. For a $((n,L,d))$ code which is
diagonal in graph state basis and encoding $L$ states into $n$
qubits with distance $d$, the joint output density matrix is
reduced to $2^n$ blocks, each is a $L\times L$ matrix. For a
stabilizer code $[[n,l,d]]$ input which is diagonal in graph state
basis and with an equal prior probability for all $2^l$ encoded
states, each block of joint output density matrix of a
depolarizing channel can be further diagonalized with the Hadamard
matrix. The eigenvalues are obtained in closed form. ''$n_1$ in
$n_2"$ concatenated repetition codes are used to illustrate the
details.

Funding by the National Natural Science Foundation of China (Grant No.
60972071), Zhejiang Province Science and Technology Project (Grant No.
2009C31060) are gratefully acknowledged.


\begin{thebibliography}{99}
\bibitem{Lloyd}  S. Lloyd, Phys. Rev. A 55, 1613 (1997).

\bibitem{Barnum1}  H. Barnum, M. A. Nielsen, and B. Schumacher, Phys. Rev. A
\textbf{57}, 4153 (1998).

\bibitem{Barnum2}  H. Barnum, E. Knill, and M. A. Nielsen, IEEE Trans. Inf.
Theory \textbf{46}, 1317 (2000).

\bibitem{Devetak}  I. Devetak, IEEE Trans. Inf. Theory \textbf{51}, 44
(2005),

\bibitem{Shor}  P. Shor and J. Smolin, arXiv quant-ph/9604006.

\bibitem{DiVincenzo}  D. DiVincenzo, P. Shor, and J. Smolin, Phys. Rev. A
\textbf{57}, 830 (1998).

\bibitem{Smith}  G. Smith and J. Smolin, Phys. Rev. Lett., \textbf{98},
030501 (2007).

\bibitem{Fern}  J. Fern and K. B. Whaley, Phys. Rev. A \textbf{78,}062335
(2008).

\bibitem{Hein1}  M. Hein, J. Eisert, and H. J. Briegel, Phys. Rev. A \textbf{%
69},062311 (2004).

\bibitem{Sch}  D. Schlingemann and R. F. Werner, Phys. Rev. A \textbf{65},
012308 (2002).

\bibitem{Raus}  R. Raussendorf and H. J. Briegel, Phys. Rev. Lett. \textbf{86%
}, 5188 (2001).

\bibitem{Walther}  P.Walther, et.al., Nature \textbf{434},(2005) 169 ; C.Y.
Lu, et.al., Nature Physics \textbf{3}, 91 (2007) .

\bibitem{Acin}  D. Cavalcanti, R. Chaves, L. Aolita, L. Davidovich, A. Acin,
Phys. Rev. Lett. \textbf{103}, 030502 (2009)
\end{thebibliography}
\end{document}